\documentclass[sigconf,authorversion,nonacm]{acmart}
\AtBeginDocument{%
  \providecommand\BibTeX{{%
    \normalfont B\kern-0.5em{\scshape i\kern-0.25em b}\kern-0.8em\TeX}}}

\acmConference[SC23]{Sustainable Supercomputing Workshop, SC23}
\acmBooktitle{Sustainable Supercomputing Workshop}

\begin{document}

\setlength{\belowcaptionskip}{-10pt}
\captionsetup{belowskip=-10pt}

\settopmatter{printacmref=false}
\title{Evaluating Total Environmental Impact for a Computing Infrastructure}

\author{Adrian Jackson}
\email{a.jackson@epcc.ed.ac.uk}
\orcid{0000-0003-0073-682X}
\affiliation{%
  \institution{EPCC, The University of Edinburgh}
  \streetaddress{Bayes Cetnre}
  \country{UK}
}

\author{Jon Hays}
\email{j.hays@qmul.ac.uk}
\affiliation{%
  \institution{Queen Mary University of London}
  \country{UK}
}

\author{Alex Owen}
\email{r.a.owen@qmul.ac.uk}
\affiliation{%
  \institution{Queen Mary University of London}
 \country{UK}
}

\author{Nicholas Walton}
\email{naw@ast.cam.ac.uk}
\affiliation{%
  \institution{Institute of Astronomy, University of Cambridge}
  \streetaddress{University of Cambridge}
  \country{UK}
}

\author{Alison Packer}
\email{alison.packer@stfc.ac.uk}
\affiliation{%
  \institution{Scientific Computing, STFC}
  \streetaddress{Rutherford Appleton Laboratory}
  \country{UK}
}

\author{Anish Mudaraddi}
\email{anish.mudaraddi@stfc.ac.uk}
\affiliation{%
  \institution{Scientific Computing, STFC}
  \streetaddress{Rutherford Appleton Laboratory}
  \country{UK}
}

\renewcommand{\shortauthors}{Jackson et al.}

\begin{abstract}
In this paper we outline the results of a project to evaluate the total climate/carbon impact of a digital research infrastructure for a defined snapshot period. We outline the carbon model used to calculate the impact and the data collected to quantify that impact for a defined set of resources. We discuss the variation in potential impact across both the active and embodied carbon for computing hardware and produce a range of estimates on the amount of carbon equivalent climate impact for the snapshot period.
\end{abstract}



\maketitle

\section{Introduction}
Moving towards Net-Zero for digital research infrastructures (DRIs), i.e. providing DRIs that do not have significant impacts on the climate or environment, requires robust information to enable good decision making around infrastructure procurement and provisioning. This requires understanding the full carbon costs or climate impacts associated with operating, maintaining, and using the infrastructure, going beyond accounting for the electricity and cooling required for operations of any service, and including the full chain of costs embodied in the infrastructure. 

In this short paper we outline the work done during the IRISCAST project~\cite{IRISCAST_Initial} to evaluate the full lifecycle climate emissions associated with an active DRI, both by cataloguing the resources that compose the DRI and by measuring energy consumption for a defined period of the operation of the DRI. To convert the collected data into impact on the climate of the DRI we have developed a carbon model to produced an overall figure for the climate impact of a 24 hour period (a snapshot) of operating the IRIS DRI.

During this process we have identified many areas where data is either incomplete or of variable quality, signalling that much more work is required to properly quantify the climate impact of DRIs, such as High Performance Computing systems (HPC). Nevertheless, this initial work, coupled with other available data, lets us start to build a picture of where the majority of climate impacts are likely to originate for DRIs and thereby let us start to address these areas to reduce the overall climate impact of DRI technologies whilst maximising the benefits such infrastructure provides.

For the rest of the paper, we will introduce the IRIS DRI, briefly discuss the IRISCAST approach, outline the carbon model we have designed, and then discuss the results of monitoring and evaluating the DRI for a 24 hour period to enable quantifying the climate impact of such a system. We finish with a discussion of the implications of this work and future research that could improve the accuracy of such measurement approaches.

\section{IRIS}
IRIS~\footnote{\url{https://www.iris.ac.uk/}} is a collaboration of computing and data storage providers that deliver a DRI supporting research in areas such as particle physics, nuclear physics, space science and astronomy. The hardware is being used for projects such as the Square Kilometre Array and Deep Underground Neutrino Experiment. Table~\ref{tab:iris-hardware} summarises  the hardware from IRIS that was includes in our snapshot experiment.

\begin{table}
  \caption{Summary of the IRIS hardware included in the project}
  \label{tab:iris-hardware}
  \begin{tabular}{p{2in}p{1.1in}}
    \toprule
    Site&Hardware\\
    \midrule
     Queen Mary University of London (QMUL) & 118 CPU nodes \\
     Cambridge University (CAM) & 60 CPU nodes \\
     Durham University (DUR) & 808 CPU nodes \\ 
     & 64 storage nodes \\
     Rutherford Appleton \newline Laboratory (STFC) & 699 CPU nodes \newline (SCARF HPC system) \\ 
     &  651 CPU nodes  \newline (STFC Cloud) \\
     &  105 storage nodes \\
     Imperial College London (IMP) &  241 CPU nodes \\
  \bottomrule
\end{tabular}
\end{table}

For the IRIS hardware under consideration we were able to collect power and/or energy data from the systems, to allow us to calculate the overall energy used by the DRI for a snapshot period. Varying methods were used to collect this data, from monitoring of the bulk power meters used to record power usage for whole machine rooms, through to rack level monitoring systems, and down to collecting on-node power measurement using software tools such as IPMI\footnote{\url{https://en.wikipedia.org/wiki/Intelligent_Platform_Management_Interface}}.

\section{IRISCAST}
The goal of the IRISCAST project was to attempt to quantify the climate impacts of IRIS, with an aim of collecting sufficient reliable data to enable future decision making about computing resource procurement and operation to incorporate potential climate impacts. With such data it should be possible to reduce the climate impacts of future DRI, either through procurement of resources that have a lower climate impact, or by operating the DRI in a way that reduces climate impacts, thereby helping with the move towards the goal of net zero digital infrastructure. 

%However, we should acknowledge that purely through operation and procurement of DRI it is not possible to reach a net zero position (i.e. have a DRI that has no climate impacts). Fully net zero DRI will require other mechanisms, such as carbon capture, local heating schemes, or climate offsetting. This does not preclude attempts to reduce the climate impact of DRIs, and the first stage of doing that is being able to properly quantify the impacts of current systems. It will also be useful to enable evaluating the various “orders of magnitude” for the carbon model constituents for representative DRIs, thereby allowing effort to be focussed on the most costly parts of the DRI when it comes to reducing climate impacts of such systems.

\section{Carbon Model}

For this work we identify two main data sources of interest:
\begin{enumerate}
\item Active (also known as operational) energy usage
\item Embodied (also known as embedded or sunk cost) carbon cost
\end{enumerate}
%Whilst these are relatively straight forward to articulate, there are lots of subtleties in how these values are defined, gathered, and evaluated, that can complicate or change the overall carbon emissions model. One of the first global considerations is the time period or duration considered for the carbon model. 

Active energy usage needs to be associated with a particular time period as, for instance, the carbon mix of energy provided or the energy cost of cooling operations will vary depending on the period considered. Energy usage of a DRI is also likely to be variable depending on the load/usage of the system. These factors mean that collecting data on, and accurately calculating, an energy consumption for the period of interest will provide the most accurate active energy usage estimation for a DRI.

However, the embodied carbon associated with DRI resources is generally a fixed cost (unless additional infrastructure or hardware is added during the period under consideration), incurred at manufacture, installation, and decommissioning. Therefore, this carbon cost needs to be apportioned to the DRI via some mechanism that will account for this embodied cost, but at a rate that is justifiable for the period the carbon emissions model is being used to evaluate. 

Given that the facilities hosting DRIs and DRI components themselves will have a range of lifetimes, this also needs to be factored into the calculation to ensure that if the full lifetime of the DRI was evaluated the total carbon cost was realised, whilst ensuring that partial lifetime evaluations (like the snapshot) have an accurate representation of embodied carbon cost as well.

\subsection{Active Energy}

The active energy usage for a DRI consists of all the energy used to run the DRI system for the period of interest. One of the challenges for this metric is deciding and defining which resources are included in the DRI, apportioning the percentage of resources shared by the DRI and other infrastructure, and defining the scope of resources.  We consider the following primary resources as active energy components for DRI style infrastructure: Compute nodes, Login nodes, Storage nodes, Service nodes, and Network.

Infrastructure that supports the DRI and contributes to the active energy usage are as follows: Cooling systems for the DRI resources and buildings hosting the DRI resources, Power distribution units and transformers supplying DRI resources and infrastructure, Uninterruptable power supply (UPS) resources supporting DRI systems, Facility electricity usage, such as lighting, fire and security systems, as well as other ancillary systems within the data centre/infrastructure hosting the DRI resources.

%Whilst it is generally straightforward to identify those components that constitute the DRI, defining the boundaries of all DRI system contributions may not be as straightforward. For instance, energy costs may be variable depending on the usage of the systems as a whole, i.e. if there is a shared cooling system and another service (not the DRI under consideration) is placing a higher load on it than the DRI is it is likely the cooling system will be consuming proportionally more energy than at lower usage conditions, which could impact the amount of energy being assigned to the DRI usage. There may also be effects from external factors, such as the weather or power distribution available at different points during the year. 

%There is no perfect way to assign such costs and to mitigate entirely for areas where resources are shared between different systems, however, what we can do is undertake an “order of magnitude” assessment of these costs to evaluate how big an impact such variability is likely to have on the overall results of the carbon model. If, for instance, it becomes evident that cooling system energy usage is <10\% of the overall DRI energy usage, and cooling system energy usage only varies itself by ±10\% across the full range of system loads, then it can be decided that such variability would have negligible impact on the overall validity of the carbon model.

\subsection{Embodied climate impacts}

For the carbon model to be accurate, we also need to consider the embodied (already created) emissions of the resources used for, or supporting the DRI, i.e. those emissions related to the manufacture, installation, and decommissioning of the DRI and associated facilities. As with the active energy costs, these can broadly be split into the primary resources the DRI is composed of, i.e. the carbon cost of manufacturing the nodes, storage, and network equipment that constitute the DRI, and the infrastructure used to host and operate the DRI, i.e. the buildings the DRI components are housed in, the cooling and power distribution infrastructure, and other associated fixed infrastructure.

%As before, these two different categories may have varying applicability or fractional assignment to the DRI in question, and also may have varying life spans. Data centre facilities are likely to house multiple systems, and also have a lifespan significantly longer than a single DRI. Individual DRI components may have varying life spans (i.e. compute nodes may be replaced more quickly than storage nodes, data tapes may have to be replaced within a specific period, etc…), DRI equipment may be added to or changed over time, and overall resources included in a DRI may change over time (as sites are added or removed).
%Facilities such as data centres, cooling systems, and power distribution systems will also have life spans, be upgraded, expanded, or changed over time, and may support a range of facilities during operations. The scale of individual resources or infrastructure may also have an impact on how efficient they are (where efficiency can be defined as the ratio of embodied carbon cost to equipment supported).

\subsection{Model}

As defined in the previous sections, the overall carbon model has the following form:
\begin{equation}
C_t^p=C_a^p+C_e^p
\end{equation}
where $C_t^p$ is the total carbon (or carbon equivalent) usage of the DRI for the time period $p$, $C_a^p$ is the active carbon usage for the same period across all DRI resources, and $C_e^p$ is the embodied carbon calculated for the DRI for that period. %Whilst simple in formulation, the complexity comes when deciding what is included $C_a^p$ and $C_e^p$ and how they are derived.

The active carbon can be considered as simply a sum of all the carbon created by the energy used by the active components of the DRI during $p$. Given the components we have already identified within earlier, we can express this as:
\begin{equation}
    C_a^p=\sum\limits_{1}^{node}\sum\limits_{1}^{p}C_{anode} +  \sum\limits_{1}^{networks}\sum\limits_{1}^{p}C_{anetwork} + \sum\limits_{1}^{facilities}\sum\limits_{1}^{p}C_{afacilities}
\end{equation}
Where all these carbon calculation take the following form:

\begin{equation}
C_ax^p=E_x^p \times CM_e^p
\end{equation}
with $C_ax^p$ being the active energy carbon impact of item $x$, $E_x^p$ is the energy used by item $x$ during the period $p$, and $CM_e^p$ is the carbon intensity factor (factor to convert the energy used into carbon equivalent units) for the electricity used. $CM_e^p$ can be derived from the electricity/energy supply mix used by item $x$ for period $p$.

For the individual elements, $E_{nodes}^p$ represents the sum of all the energy used by all the nodes involved in the DRI. We assume that nodes are fully assigned to the DRI, %although the model could accommodate partial resource assignment (nodes only partially used by the DRI), provided the embodied carbon model also reflects this partial assignment.
$E_{network}^p$ represents the energy used by the network to connect to the nodes and used for communication between nodes and any external network required by the DRI. $E_{facilities}^p$ can be further split into the following:
    \begin{itemize}
	\item $E_{cooling}^p$: The energy consumed by the cooling systems utilised by the DRI
	\item $E_{power}^p$: The energy consumed by the power distribution (i.e. transformer losses) and uninterruptable power supply (UPS) systems supporting the DRI
	\item $E_{facility}^p$: The energy used by the wider facility(s) supporting the DRI (building lights, local computers, heating, etc. for the data centre(s) hosting the DRI hardware)
    \end{itemize}

The embodied carbon costs can be modelled in the same way as the active energy: 

\begin{equation}
    C_e^p=\sum\limits_{1}^{node}\sum\limits_{1}^{p}C_{enode} +  \sum\limits_{1}^{networks}\sum\limits_{1}^{p}C_{enetwork} + \sum\limits_{1}^{facilities}\sum\limits_{1}^{p}C_{efacilities}
\end{equation}

For embodied carbon, $C_{enod}e$ is the carbon emitted in creating, delivering, installing, and decommissioning a given node, likewise for network and facility components in the system (including buildings and infrastructure as appropriate). Embodied carbon is not dependent on the time period under consideration, which means there is a requirement to spread these emissions across the period being evaluated. 

For instance, if a compute node is responsible for 5kg of embodied carbon, and the node will be used for 5 years, we could consider that to be 1kg of carbon per year. If the period under consideration for the carbon model is 6 months, then we could consider that 500 grams of carbon be a sensible amount to assign for the carbon model for this specific evaluation.

%When considering facilities, such as buildings, or power/cooling infrastructure, the same is true, with buildings generally having considerably longer life spans than both the DRI and any particular carbon model snapshot that could be considered.  Therefore, the carbon model should be transparent about the assignment of resources to the model, and ensure the assignment value is consistent between the active and embodied evaluation parts of the model.

\section{Evaluation}

To collect the active power for the snapshot period, a number of different methods were used, depending on what was available at the specific systems under consideration. These included collecting facility level data (i.e. power data at the level of a machine room or facility), individual power distribution unit readings for the hardware under consideration, and using on-node software/hardware approaches such as IPMI and Turbostat\footnote{\url{https://www.linux.org/docs/man8/turbostat.html}}.

The data collected is presented in Table~\ref{tab:iris-energy}, from which we can see there is some variation between different power/energy measurement approaches, but the total energy used was just under 19,000 kWh. The variation between measurement approaches does demonstrate that care is needed in collecting this data and potentially adjusting measurements to get an accurate overall measurement for a facility (i.e. adjusting in-node energy/power data to reflect the overheads that are not being collected). For example, we can see that the Turbostat readings for QMUL in Table~\ref{tab:iris-energy} are $~5\%$ lower than the IPMI figures, which are $~1.5\%$ lower than the PDU figures.

\begin{table}
  \caption{Active energy measured for the snapshot period (kWh)}
  \label{tab:iris-energy}
  \begin{tabular}{lccccc}
    \toprule
    Site&Facility&PDU&IPMI&Turbostat&Nodes \\
    \midrule
    QMUL 	& 1299	& 1299  &	1279 	& 1214	& 118 \\
    CAM 	& 261	& -	& 261	& - &	59 \\
    DUR 	& 8154 &	8154 &	6267	& -	& 876 \\ 
    STFC CLOUD &	3831	& - &	3831	& - &	721  \\
    STFC SCARF &	 4271 & 	4271 & 	3292 &	 - &	571  \\
    IMP 	& 944 & - & 944 & - &	117 \\ 
    Total  &	18760	 & & & & \\
  \bottomrule
\end{tabular}
\end{table}

\begin{figure}[h]
  \centering
  \includegraphics[width=\linewidth]{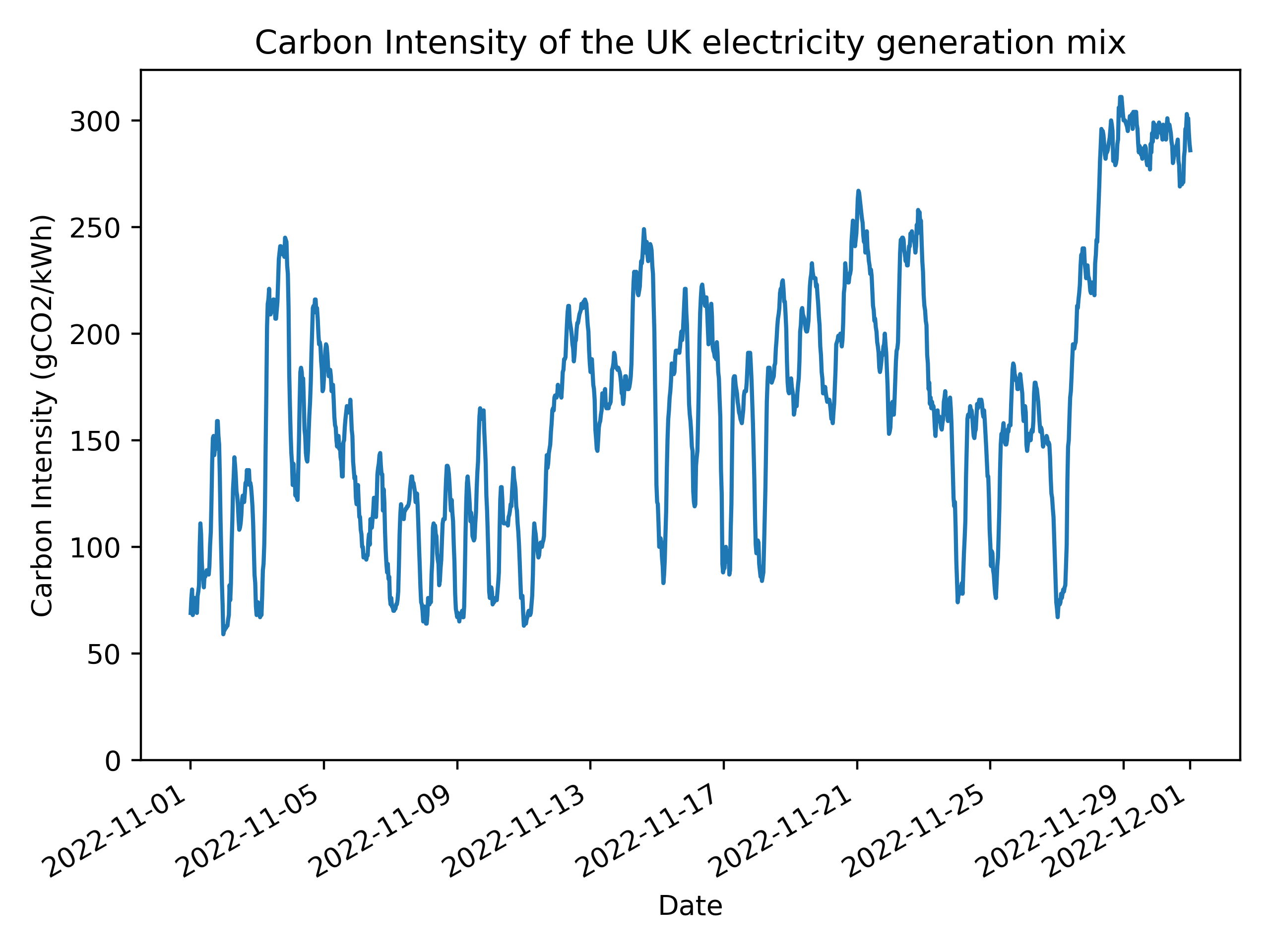}
  \caption{UK electricity generation carbon intensity for the month of November 2022}
  \Description{A graph of the carbon intensity of UK electricity generation for the month of November 2022}
  \label{fig:carbon_intensity}
\end{figure}

To get an assessment of the climate impact of this energy usage we convert the kWh used into a measure of carbon equivalent units. We have evaluated the carbon intensity of the UK electrical generation mix around the period our tests were undertaken using data from the Carbon Intensity website\footnote{\url{https://carbonintensity.org.uk/}}. Figure~\ref{fig:carbon_intensity} is the average carbon intensity of the UK electricity provision over the month of November 2022 (when the snapshot was undertaken).

\begin{table*}
  \caption{Active Carbon Estimates ($kgCO_2$)}
  \label{tab:active_carbon}
  \begin{tabular}{p{1in}p{0.4in}p{0.4in}p{0.4in}p{0.4in}p{0.4in}p{0.4in}p{0.4in}p{0.4in}p{0.4in}}
    \toprule
    \textbf{Metric}&\multicolumn{3}{c}{\textbf{Low}}&\multicolumn{3}{c}{\textbf{Medium}}&\multicolumn{3}{c}{\textbf{High}} \\
    \midrule
    Active Energy \newline Carbon 	& \multicolumn{3}{c}{969}	& \multicolumn{3}{c}{3391}	& \multicolumn{3}{c}{5814}\\
    \midrule
    \textbf{PUE Estimate}	& \textbf{Low}	&  \textbf{Medium} &	 \textbf{High} &	 \textbf{Low}  &	 \textbf{Medium}	&  \textbf{High} &	 \textbf{Low} 	&  \textbf{Medium}	& \textbf{High} \\
    \midrule
    Active Energy \newline Carbon including \newline Facilities  &	1066 &	1260	&  1550	 & 3731	& 4409	& 5426	
& 6395	&7558	&9302 \\
  \bottomrule
\end{tabular}
\end{table*}

As Figure~\ref{fig:carbon_intensity} demonstrates, there is significant variability in the generation carbon intensity. As such, we are selecting three reference values to utilise in the evaluation of carbon generated by the active energy for IRIS: A \textbf{Low} value of 50, a \textbf{Medium} value of 175, and a \textbf{High} value of 300$gCO_2 \slash kWh$. These gives values of 969, 3391, and 5814$kgCO_2$ respectively for the energy used to run the compute hardware in the DRI.

None of the facilities under consideration were able to provide measures for the electricity used for cooling or other infrastructure, so we estimate these using a range of PUE factors (PUE: Power Usage Effectiveness, quantifies how much extra electricity is required to run infrastructure above the electricity consumed by the compute hardware itself). For the sake of this evaluation we use a \textbf{Low} value of PUE of 1.1, a \textbf{Medium} value of 1.3, and a \textbf{High} value of 1.5. If we combine these with the carbon equivalent production we have already calculated for the different carbon intensities we get a range of potential active carbon production outlined in Table~\ref{tab:active_carbon}.

Information about the compute hardware monitored during the snapshot period was provided in the inventories from each facility.  Manufacturers are starting to provide embodied carbon estimates for hardware they are selling or have sold in the past. Examples of such assessments are available\footnote{\url{https://i.dell.com/sites/content/corporate/corp-comm/en/Documents/dell-server-carbon-footprint-whitepaper.pdf}}\footnote{\url{https://www.fujitsu.com/global/documents/about/environment/Life\%20cycle\%20analyses\%20of\%20Fujitsu\%20Desktop\%20ESPRIMO\%20P9010\%20June\%202021.pdf}}, and include raw materials carbon impact, assembly carbon costs (the construction of the hardware), and transportation costs to get all the parts assembled and sent to the end user. These estimates are provided as $kgCO_2$ costs, with the conversion factor from the raw energy or material usage to actual emissions pre-calculated.

Using a range of such information we came up with two embodied carbon estimates for a notional compute node, 400 and 1100$kgCO_2$. These encapsulate the high and low ends of the estimates we have seen for server node embodied carbon, and let us calculate upper and lower bounds on the embodied carbon of the DRI. Using these, and then a varying lifespan of the hardware, we can create some estimates of the embodied carbon cost for the DRI for a 24 hour snapshot period, as outlined in Table~\ref{tab:embodied_carbon}.

\begin{table}
  \caption{Active energy measured for the snapshot period (kWh)}
  \label{tab:embodied_carbon}
  \begin{tabular}{p{0.5in}p{0.5in}p{0.5in}p{0.5in}p{0.5in}}
    \toprule
    &\multicolumn{4}{p{2in}}{Embodied carbon Estimate ({$kgCO_2$})}\\
    Server & 400 & 1100 & 400 & 1100 \\
    \midrule
    Server Lifespan \newline(years)	& \multicolumn{2}{p{1in}}{Embodied carbon \newline({$kgCO_2$} per 24 hours per server)}& \multicolumn{2}{p{1in}}{Snapshot Embodied carbon \newline({$kgCO_2$})} \\
    \midrule
    3 	& 0.36	& 1.00	& 876	& 2409 \\
    4 	& 0.27 &	0.75 &	657	& 1806 \\ 
    5 &	0.22	& 0.61 &	526	& 1445  \\
    6 &	 0.18 & 0.50 & 	438 &	1204   \\
    7 	& 0.16 & 0.43 & 375 & 1032 \\ 
  \bottomrule
\end{tabular}
\end{table}

\section{Summary and Future Work}

A more thorough evaluation of the embodied carbon of the DRI under consideration would require getting specific data for each type of node being used and using that instead of our estimated figures. Likewise, we should also collect data from the cooling and power hardware to obtain actual PUE measurements rather than estimates. We have also left out data for the embodied carbon associated with the data centre infrastructure (building, cooling hardware, etc...) because of space constraints for this paper. 

All these inputs are required to gain a more accurate carbon estimate for the DRI, but what we have done has given us an initial evaluation of the magnitude of different components of the climate impact for IRIS. We can see that we have a range of estimates for the embodied carbon for a 24 hour period of operation of IRIS of between 375 and 2409$kgCO_2$, and a range of active carbon for the same period of 1066 to 9302$kgCO_2$. This demonstrates that the embodied carbon is generally a much smaller percentage of the overall impact for most scenarios outlined, making the active carbon a higher priority for initial efforts in reducing the climate impact of DRIs.

The positive aspect of this conclusion is that there is already significant work underway to decarbonise energy generation around the global, meaning the overall climate impacts of active part of DRIs is set to decrease over time. However, this does mean the embodied carbon will come to dominate the climate impact of such systems, meaning that a focus on decarbonising manufacture and supply of computing hardware will become important in the near future.

Furthermore, whilst there are scenarios where the energy supply mix will have zero direct carbon associated with it (i.e. if the energy is generated from fully renewable sources), we should be aware that even renewable energy sources have carbon emissions associated with them, either from the embodied costs associated with manufacturing and installing the energy sources, or from the emissions associated with operating those resources (for instance, gases released from lakes created for hydro-electric power). It is also worth considering that until the entire energy network is fully renewable, using zero carbon energy for DRI operations may simply be displacing other activities onto more carbon heavy energy sources.

We should also be cognisant that this assessment of the climate impact of a DRI does not consider what the DRI was actually being used for, how efficiently jobs were running on the DRI, or any other usage questions around DRI and computing resources in general. It does not consider the positive impact work on DRIs can have on climate change and environmental issues (i.e. by developing and designing net zero technologies).  Furthermore, it is also worth considering the magnitude of the carbon impact of this DRI to other activities, for instance carbon emissions associated with flying can be estimated at 92$kgCO_2$ per passenger per hour\footnote{\url{https://www.carbonindependent.org/22.html}} for a representative jet engine powered aircraft. If we flew a person for the 24-hour period of our snapshot their associated carbon emissions would be 2208$kgCO_2$. Putting this into perspective, the estimated carbon emissions for IRIS of our snapshot period are between 1 and 4 of these passenger journeys (or, more realistically, equivalent to 1 to 4 people travelling on a return 12-hour flight).

\begin{acks}
The IRISCAST project was supported by the NetZero Scoping project~\cite{netzero_final_report}, which was funded  by the UKRI Digital Research Programme on grant NERC (NE/W007134/1).
\end{acks}
\bibliographystyle{ACM-Reference-Format}
\bibliography{sc23_sustain_workshop_iriscast}

\end{document}